\documentclass[
    aps,
    pre,
    reprint,
    superscriptaddress,
    longbibliography,
    amsmath,
    amssymb,
    floatfix
]{revtex4-2}

\usepackage[T1]{fontenc}
\usepackage{graphicx}
\usepackage{bm}
\usepackage{siunitx}
\usepackage[version=3]{mhchem}
\usepackage{xcolor}
\usepackage{microtype}
\usepackage[normalem]{ulem}
\usepackage{soul}
\usepackage{hyperref}

\begin{document}

\title{Frequency-Controlled Penetration and Scattering
of Driven Particles at Obstacle Lattices}

\author{Galor Geva}
\affiliation{Department of Theoretical Condensed Matter
Physics, Universidad Autonoma de Madrid,
Madrid 28049, Spain}
\affiliation{Condensed Matter Physics Center (IFIMAC),
Universidad Autonoma de Madrid, Madrid 28049, Spain}
\affiliation{Instituto Nicolas Cabrera,
Universidad Autonoma de Madrid, Madrid 28049, Spain}

\author{Arin Escobar Ortiz}
\affiliation{Department of Theoretical Condensed Matter
Physics, Universidad Autonoma de Madrid,
Madrid 28049, Spain}
\affiliation{Condensed Matter Physics Center (IFIMAC),
Universidad Autonoma de Madrid, Madrid 28049, Spain}
\affiliation{Instituto Nicolas Cabrera,
Universidad Autonoma de Madrid, Madrid 28049, Spain}

\author{Laura R. Arriaga}
\email{laura.rodriguezarriaga@uam.es}
\affiliation{Department of Theoretical Condensed Matter
Physics, Universidad Autonoma de Madrid,
Madrid 28049, Spain}
\affiliation{Condensed Matter Physics Center (IFIMAC),
Universidad Autonoma de Madrid, Madrid 28049, Spain}
\affiliation{Instituto Nicolas Cabrera,
Universidad Autonoma de Madrid, Madrid 28049, Spain}

\author{Juan L. Aragones}
\email{juan.aragones@uam.es}
\affiliation{Department of Theoretical Condensed Matter
Physics, Universidad Autonoma de Madrid,
Madrid 28049, Spain}
\affiliation{Condensed Matter Physics Center (IFIMAC),
Universidad Autonoma de Madrid, Madrid 28049, Spain}
\affiliation{Instituto Nicolas Cabrera,
Universidad Autonoma de Madrid, Madrid 28049, Spain}

\begin{abstract}
The transport of driven particles through structured media is strongly influenced by hydrodynamic interactions with nearby boundaries. While the role of media geometry in shaping these interactions has been widely explored, less attention has been given to how the driving frequency modifies particle transport through a fixed environment. Here, we study a rotating magnetic particle approaching a square lattice of cylindrical obstacles, varying both the rotation frequency and the obstacle spacing. At low frequencies, the particle penetrates the lattice, whereas at higher frequencies it is scattered and slides along the lattice edge. We show that this transition originates from the hydrodynamic interaction with individual finite-height obstacles, which generates a vertical lift that competes with the frequency-independent buoyant weight. The resulting vertical displacement weakens the hydrodynamic coupling with the substrate, leading to a scattering distance that increases as $r_{\mathrm{sc}}\sim f^{1/2}$. When this distance becomes sufficiently large relative to the spacing between neighboring obstacles, the lattice edge acts as a hydrodynamic barrier that prevents penetration. These results show how driving frequency and geometrical confinement jointly determine the pathways of driven particles through structured environments, even in the Stokes regime
\end{abstract}

\maketitle

\section{Introduction}\label{sec1}
At low Reynolds numbers, the motion of driven particles through structured environments is shaped by hydrodynamic interactions with nearby boundaries and obstacles. These interactions couple the particle motion to the surrounding geometry through the flow and pressure fields established in the confined fluid, making transport through porous or crowded media sensitive not only to particle properties but also to the structure of the environment~\cite{witten2020review}. The role of obstacle arrangement is well documented: periodic post arrays, microchannels, and funnel geometries can guide, trap, rectify, or enhance particle motion by modifying the surrounding flow and the hydrodynamic interactions with nearby boundaries~\cite{dehkharghani2023self,van2022role,makarchuk2019enhanced,cisneros2006reversal}. These effects have been exploited to separate particles by size, shape, and motility in systems ranging from synthetic colloids to bacteria and sperm cells~\cite{shi2009continuous,huang2004continuous,yamada2004pinched,di2007continuous,khan2024acoustofluidic,petersson2004separation,galajda2007wall,zhou2024ai,chinnasamy2018guidance,yazdan2022sperm,nath2023microfluidic,misko2023selecting}. In these approaches, the obstacle geometry typically provides the main control parameter. Less is known about how the driving frequency or activity modifies the hydrodynamic interaction with the environment, and whether varying this frequency alone, at fixed obstacle geometry, can produce qualitatively different transport regimes.

Magnetic microrollers, particles driven into rotation by a rotating magnetic field, are well-suited to address this question. Near a substrate, the broken symmetry of the rotational flow couples particle rotation to translation~\cite{tierno2008magnetically,sing2010controlled,tasci2016surface}. The rotation frequency controls the particle velocity and associated hydrodynamic stresses while leaving the geometry unchanged, making it a convenient control parameter for particle transport. When a microroller approaches an obstacle, the hydrodynamic interaction with the obstacle generates a pressure distribution that deflects the particle and can lift it away from the substrate. The resulting trajectory reflects a competition between substrate-induced propulsion and obstacle-induced pressure. Prior work on this interaction showed that when the obstacle is much larger than the particle, it effectively behaves as an unbounded wall, and the particle becomes hydrodynamically trapped at a well-defined orbit radius, escaping only through thermal fluctuations~\cite{van_der_wee_simple_2023}. More recent experiments and simulations have further shown that the driving frequency can give rise to distinct dynamical behaviors through hydrodynamic interactions alone~\cite{magrinya_rolling_2025,geva2025rolling}. However, the role of finite obstacle size remains less understood, particularly whether finite-height obstacles generate vertical hydrodynamic forces absent in the large-obstacle limit, thereby modifying the particle-substrate separation and the shear stresses that drive its motion.

\begin{figure}[t]
    \centering
\includegraphics[width=0.95\linewidth]{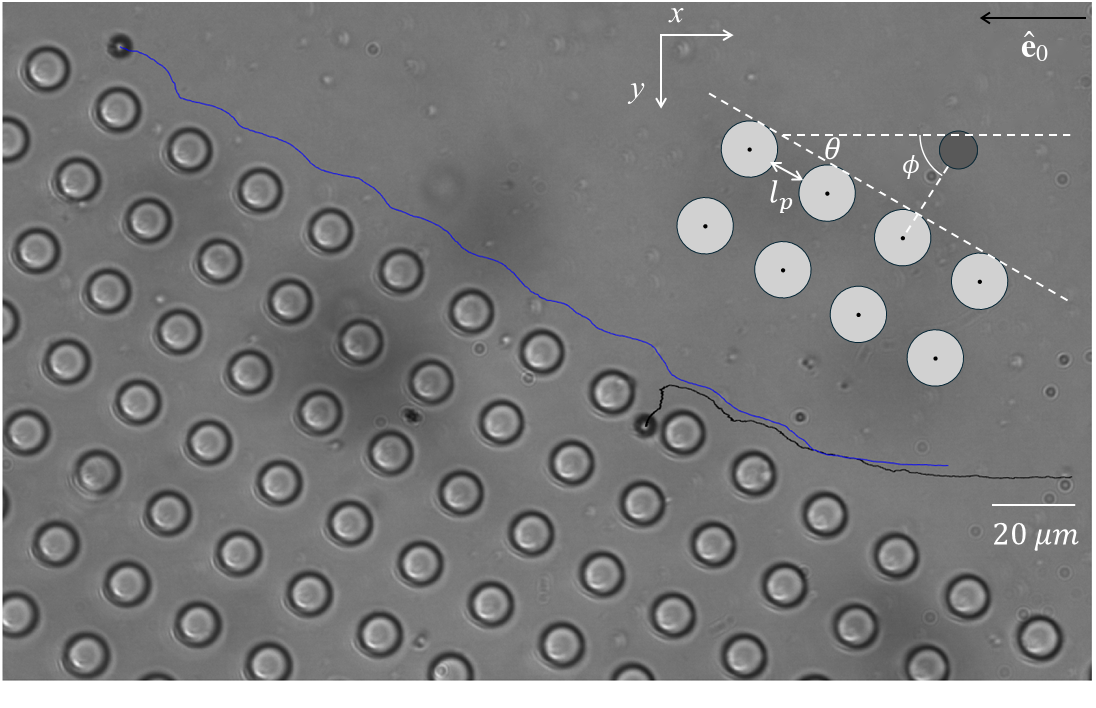}
    \caption{Trajectories of particles rotated toward the lattice, at frequencies of 10 (black) and 40 (blue) Hz. Inset shows the incidence angle $\theta$, the angular coordinate $\phi$, and the edge-to-edge lattice spacing $l_p$.}
    \label{fig1_new}
\end{figure}

In this work, we address this question using magnetically driven microrollers approaching a square array of cylindrical obstacles (Fig.~\ref{fig1_new}). The obstacle height and radius are comparable to the particle size, placing the system outside the large-obstacle limit considered previously. 
For an isolated obstacle, the finite obstacle height allows the rotational flow generated by the particle to escape asymmetrically over the obstacle top, producing a vertical hydrodynamic lift that is absent in the large-obstacle limit~\cite{geva2025rolling}. This lift increases with rotation frequency, displacing the particle farther from the substrate and weakening the shear stresses that sustain its rolling motion. The resulting modification of the balance between substrate-induced shear and obstacle-induced pressure shifts the scattering trajectory to larger particle--obstacle separations, while the minimum approach distance remains nearly unchanged. In the full lattice, the frequency-dependent scattering distance determines whether the particle penetrates the array or is redirected along its boundary. At low frequencies, particles pass between neighboring obstacles, whereas at high frequencies the scattering regions of adjacent obstacles overlap and collectively form a hydrodynamic barrier. For oblique incidence, scattered particles slide along the lattice edge, with trajectories crossing over from smooth wall-like guidance to discrete obstacle-to-obstacle motion as the lattice spacing increases. When several particles slide simultaneously along the boundary, hydrodynamic interactions between them lead to faster collective motion than for isolated rollers. These results show how finite obstacle size, driving frequency and lattice spacing jointly determine transport through structured media.

\section{Materials and Methods}

\subsection{Fabrication of the PDMS substrates}
The PDMS substrates were prepared by cast molding~\cite{xia_soft_1998} on a silicon
master wafer fabricated by photolithography (ELVESIS). The elastomer base and curing
agent of a Sylgard 184 silicone elastomer kit (Dow) were mixed at a 10:1 ratio, poured
over the master wafer, and placed in an oven at room temperature. The oven was heated
gradually over one hour to $65^\circ$C, held at $65^\circ$C for 5~h, and then allowed
to cool gradually, after which the PDMS was left overnight at room temperature.

The cured PDMS was peeled off from the master wafer, immersed in isopropanol, sonicated for
$\sim$2~min, rinsed with water, and gently dried with compressed air. To increase
wettability and reduce nonspecific interactions between the PDMS surface and the
particles, the PDMS substrate was treated with oxygen plasma for 60~s (Femto-SLS,
Diener) and immediately immersed in a 0.1~M aqueous solution of poly(ethylene glycol)
(PEG, 6~kDa, 81260, Sigma-Aldrich) for $\sim$30~min~\cite{long_polyethylene_2017}.
Finally, the substrate was rinsed thoroughly with water and gently dried with
compressed air.

\subsection{Particles and sample preparation}
We used streptavidin-functionalized magnetic particles containing magnetite,
$5~\mu\mathrm{m}$ in diameter (micromer\textsuperscript{\textregistered}-M, micromod),
coated with biotin-PEG-COOH (5~kDa, Polysciences). Samples were prepared by placing
$8~\mu\mathrm{L}$ of a dilute aqueous particle suspension on top of the PDMS
substrate, closing the sample from above with a glass coverslip, and sealing it with
a UV-cured optical adhesive (NOA 68, Norland). The particles sedimented onto the
patterned substrate prior to the experiments.

\subsection{Magnetic actuation}
Two sinusoidal signals with a phase shift of $\pi/2$ were generated using a custom
MATLAB script and converted from digital to analog via a DAQ device (Measurement
Computing USB-1208HS). The signals were amplified using 500~W amplifiers (LD Systems),
monitored with an oscilloscope (Tektronix TDS 2014B), and sent to two pairs of
Helmholtz coils aligned along the $x$ and $z$ axes, creating a magnetic field rotating
in the $xz$ plane and inducing microroller motion along the $x$ axis. The magnetic
field was $\sim$10~mT for $f \leq 20$~Hz and $\sim$15~mT for $20 < f \leq 40$~Hz, ensuring synchronous rotation in this frequency range.

\subsection{Experimental system}
The experimental system consists of magnetic microparticles, with nominal radius $a = 2.5~\mu\mathrm{m}$, rolling on a poly(dimethylsiloxane) (PDMS)
substrate patterned with cylindrical obstacles (Fig.~\ref{fig1_new}). The obstacles
have a radius of $5~\mu\mathrm{m}$ and a height of $11.5 \pm 0.5~\mu\mathrm{m}$, and
are arranged either as isolated obstacles or in square lattices with edge-to-edge
spacings $l_p = 6$, $8$, $11$ and $14~\mu\mathrm{m}$. As a limiting case, we also
used a continuous PDMS wall of the same height. The particles were driven by a
magnetic field rotating in the plane perpendicular to the substrate, over a frequency range of $f=5$--$40$~Hz. The rotating field induces a synchronous rotation of the particle, which,
through hydrodynamic coupling with the substrate, results in translation along the
free rolling direction $\hat{\mathbf{e}}_0$. With rolling velocities $v \lesssim 20~\mu\mathrm{m/s}$, the particle Reynolds number is $\mathrm{Re} = \rho_w v a/\eta \lesssim 10^{-4}$ and thus, inertia is negligible and the flow is governed by the linear Stokes equations. The lattice edge was oriented either perpendicular ($\theta = 90^\circ$) or at $\theta = 45^\circ$ relative to $\hat{\mathbf{e}}_0$ by rotating the sample relative to the plane of the rotating field.

\subsection{Imaging and particle tracking}
Images were acquired using a Nikon ECLIPSE Ts2R inverted microscope equipped with a
$40\times$ air objective (Nikon S Plan Fluor ELWD ADM 40XC, NA = 0.6) with a tunable
working distance, at a frame rate of 40 fps (BFLY-PGE-23S6M-C, FLIR). Particles and
obstacles were tracked using standard particle tracking
algorithms~\cite{crocker_methods_1996}.

\subsection{Numerical simulations}
To simulate the experimental system, we used a homemade Stokesian Dynamics \cite{Fiore_Swan_2019} code written in CUDA/C++. The code uses the Ermak--McCammon algorithm \cite{ermak_mckammon} to resolve the particle's motion, with the Rotne--Prager--Yamakawa (RPY) tensor as the mobility matrix. The RPY tensor gives the response in velocity of a sphere $i$ to a force exerted upon a sphere $j$, of equal radii $a$; with relative position $\overrightarrow{r} = \overrightarrow{r_i} - \overrightarrow{r_j}$, when immersed in a fluid of viscosity $\eta$ and in Stokes conditions ($Re = 0$). It takes the following form:
\begin{widetext}
\begin{equation}
\overrightarrow{U}_{ij} = 
\frac{1}{6 \pi \eta a} \left\{
\begin{array}{lr}
(\frac{3a}{4r} + \frac{a^{3}}{2r^{3}})\mathbf{I} + (\frac{3a}{4r} - \frac{3a^{3}}{2r^{3}})\hat{r}\hat{r}, & \text{if } i \neq j \text{ and } r \geq 2a,\\
(1 - \frac{9r}{32a})\mathbf{I} + \frac{3r}{32a}\hat{r}\hat{r}, & \text{if } i \neq j \text{ and } r < 2a,\\
 \mathbf{I}, & \text{if } i = j
\end{array}
\right\} \overrightarrow{F_j}
\end{equation}
\end{widetext}

Thermal noise was calculated using the Krylov subspace approximation ~\cite{krylov_subsp_methods}. The lubrication correction for close-range hydrodynamic interactions was not included. To resolve the effect of higher-order contributions upon the particle (such as the stresslet or the effect of the velocity gradient) we used the so-called ``raspberry model'', by which solid bodies are approximated as spherical beads connected by harmonic potentials. It has been shown that the effect of the RPY tensor upon these spherical beads approximates well the effect of higher-order contributions upon the solid body \cite{floren_raspberry_model}. We chose open boundary conditions to prevent any possible spurious hydrodynamic effects that could arise from the particle interacting with its own images in periodic boundary conditions.

The particle was simulated as a 12-bead assembly, with the beads placed as the vertices of a regular dodecahedron. The obstacles were simulated as 32-bead cylinders, with twice the diameter and 3.2 times the height of the particle. The substrate was simulated as a square bead lattice. The particle beads are connected among themselves with harmonic potentials of equilibrium length equal to the initial bead-to-bead distance and strength of 100 $k_B T$. The substrate and obstacle beads are connected among themselves in an identical manner. SM3 is a rendering of one such simulation.

To simulate the particle's rotation, a ``magnetic dipole'' $\overrightarrow{m}$ of the particle was defined as a vector connecting two diametrically opposed beads. In our case, we chose the pair with maximum and minimum $z$ in the initial configuration, so that the particle's magnetic dipole is initially parallel to the $z$ axis. A magnetic field vector $\overrightarrow{B}$ was defined as rotating in the $yz$ plane with fixed amplitude and frequency. The torque the field exerts on the particle is therefore $\overrightarrow{\tau}=\overrightarrow{B} \times \overrightarrow{m} = B_0(0, \text{ sin} (\omega t), \text{ cos} (\omega t))\times \overrightarrow{m}$. This torque is subsequently spread as forces over the 12 beads that compose the particle before carrying out the integration step.

\section{Results and Discussion}\label{sec2}

\subsection{Scattering from a single obstacle}

\begin{figure*}[t]
 \begin{center}
 \includegraphics[width=0.7\linewidth]{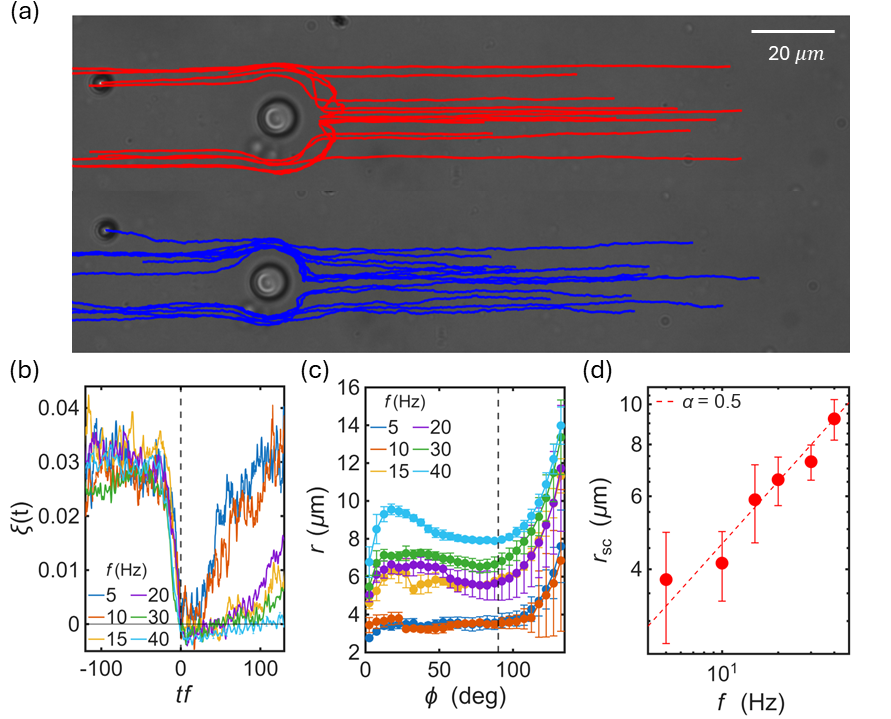}
 \end{center}
\caption{\textbf{Scattering of a driven microroller by a single obstacle.}
\textbf{(a)} Microscopy images with superimposed particle trajectories at $f = 5$~Hz (blue) and $f = 40$~Hz (red).
\textbf{(b)} Rolling parameter $\xi(t) = v(t)/(2\pi a f)$ as a function of dimensionless time $t f$. The dashed vertical line marks $t = 0$, defined as the instant the particle transiently stops before moving away from the obstacle. Negative values correspond to a brief reversal of the particle motion.
\textbf{(c)} Particle--obstacle separation $r$ as a function of the angular coordinate $\phi$ between $\hat{\mathbf{e}}_0$ and the obstacle-to-particle vector. The minimum defines $r_\mathrm{min}$; the plateau defines $r_\mathrm{sc}$.
\textbf{(d)} steady-state distance $r_\mathrm{sc}$ as a function of rotation frequency. The dashed line shows a power-law fit $r_\mathrm{sc} \sim f^{\alpha}$, with $\alpha \simeq 0.5$.
}
\label{fig2_new}
\end{figure*}

We examine the scattering of an individual microroller by a single obstacle. The particle rolls toward the obstacle along the free rolling direction $\hat{\mathbf{e}}_0$ until its motion is arrested at a minimum particle--obstacle separation $r_\mathrm{min}$. It then moves away from the obstacle and relaxes to a larger quasi-steady separation, which we define as the scattering distance $r_\mathrm{sc}$ (Fig.~\ref{fig2_new}a; see also Movies SM1 and SM2, and ESI, Fig.~S1). Once this distance is reached, the particle moves around the obstacle, completing nearly half a revolution before escaping and recovering its motion along $\hat{\mathbf{e}}_0$.

The evolution from $r_\mathrm{min}$ to $r_\mathrm{sc}$ is captured by the dimensionless rolling parameter $\xi(t) = \frac{v(t)}{2\pi a f}$, where $v(t)$ is the instantaneous particle velocity, $a$ is the microroller radius and $f$ is the imposed rotation frequency (Fig.~\ref{fig2_new}b). Far from the obstacle, $\xi(t)$ remains nearly constant, reflecting steady rolling driven by the shear stresses generated near the substrate, which exceed the opposing pressure contribution arising from the rotational flow interacting with the substrate~\cite{magrinya_rolling_2025}. As the particle approaches the obstacle, $\xi(t)$ decreases and reaches zero, which we define as $t=0$ and identify with the hydrodynamic collision point (Fig.~\ref{fig2_new}b). For $t>0$, $\xi(t)$ becomes transiently negative, indicating a brief reversal of the particle motion as the obstacle-induced pressure drives it away from $r_\mathrm{min}$ toward $r_\mathrm{sc}$. The two characteristic distances become particularly clear when the particle--obstacle separation $r$ is plotted against the angular coordinate $\phi$, defined as the angle between $\hat{\mathbf{e}}_0$ and the vector joining the obstacle center to the particle center (Fig.~\ref{fig2_new}c). The minimum of this curve defines $r_\mathrm{min}$, where forward rolling is arrested, whereas the plateau defines $r_\mathrm{sc}$.

\begin{figure}[ht!]
\begin{center}
\includegraphics[width=0.9\linewidth]{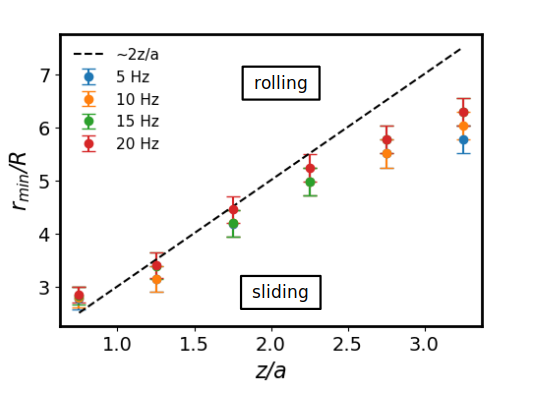}
\end{center}
\caption{\textbf{Fixed-height simulations of the minimum approach distance.} 
Minimum particle--obstacle separation $r_\mathrm{min}$, normalized by the obstacle radius $R$, as a function of the prescribed particle height $z/a$. Data at different rotation frequencies collapse onto a single trend, confirming that $r_\mathrm{min}$ is set by particle height and obstacle geometry rather than by driving frequency. The regions labeled rolling and sliding indicate the sense of the lateral force balance at each height.
}
\label{simulationfig}
\end{figure}

Comparing trajectories across rotation frequencies shows that the two characteristic distances respond differently to the driving. The minimum distance, $r_\mathrm{min}$, depends only weakly on $f$ (ESI, Fig.~S2), whereas the scattering distance, $r_\mathrm{sc}$, increases systematically, following $r_\mathrm{sc} \sim f^{1/2}$ (Fig.~\ref{fig2_new}d). This contrasting behavior indicates that $r_\mathrm{min}$ and $r_\mathrm{sc}$ are governed by different hydrodynamic balances, which we describe in terms of the shear and pressure contributions to the fluid stress acting on the particle surface. Owing to the linearity of the Stokes equations, both contributions are proportional to the imposed rotation frequency. The minimum distance $r_\mathrm{min}$ is set by a lateral balance between these two contributions: the shear force that drives rolling and the opposing pressure force that builds up in the particle--obstacle gap. The shear force scales as~\cite{goldman1967slow},
\begin{equation}
F_\mathrm{shear} \sim \eta a^2 f\, \delta(z/a),
\end{equation}
where $\eta$ is the fluid viscosity, $z$ the particle height and $\delta(z/a)$ the height-dependent lubrication coupling. In an unbounded fluid, the rotlet flow generated by the particle, $u_{\mathrm{rot}}(r) \sim a^3 f /r^2$, carries no pressure. Pressure only arises from the presence of no-slip boundaries. For a free roller, the substrate generates a pressure gradient along the rolling direction. The presence of obstacles gives rise to an additional pressure in the particle-obstacle gap that opposes the approach. Balancing this pressure gradient and viscous stresses in the Stokes equation gives $p \sim \eta u_{\mathrm{rot}}/a$. Integrating this pressure over the particle surface gives,
\begin{equation}
F_p \sim p a^2 \sim \eta a u_{\mathrm{rot}} \sim \frac{\eta a^4 f}{r^2}.
\label{pressure}
\end{equation}
Although Eq.~\ref{pressure} is an effective scaling at separation distances comparable to the particle size, both lateral forces are linear in $f$. Their ratio is therefore independent of frequency, and $r_{min}$ is set by the particle height and obstacle geometry alone. Fixed-height RPY simulations confirm this: for each prescribed height, data at different frequencies collapse onto a single curve, as shown in Fig.~\ref{simulationfig}.

The origin of $r_\mathrm{sc}$ lies in a vertical hydrodynamic lift generated by the finite obstacle height. For an obstacle much taller than the particle, roto--translational coupling could give rise to a lift, similar to how a microroller moves on a flat substrate. However, such lift occurs only at short distances~\cite{delmotte2019hydrodynamically,gao2026dynamics}. For distances comparable to the particle radius or larger, the rotational flow above and below the particle equator is distorted approximately symmetrically, and the corresponding vertical pressure contributions cancel. In our system, however, both the obstacle height and radius are comparable to the particle radius. The upper part of the rotational flow can escape over the obstacle top, whereas the flow below the particle equator remains confined between the particle, the obstacle surface, and the substrate. This asymmetry in the pressure distribution generates a net upward force on the particle. Since this lift originates from the same obstacle-induced pressure field as $F_p$~\cite{geva2025rolling,yang_microwheels_2019}, we write
\begin{equation}
F_L \sim \chi \frac{\eta a^4 f}{r^2},
\end{equation}
where $\chi$ is a dimensionless lift coefficient that quantifies the vertical component generated by the finite obstacle geometry. The lift weakly displaces the particle away from the substrate, reducing $\delta(z/a)$ and thereby weakening $F_\mathrm{shear}$, while the obstacle-induced pressure force is less strongly affected. The lateral force balance is therefore recovered at a larger particle--obstacle separation, defining $r_\mathrm{sc}$. At this separation, the lift becomes comparable to the buoyancy-corrected particle weight, $F_g = (4\pi a^3/3)(\rho_p - \rho_w)g$, giving
\begin{equation}
r_\mathrm{sc}
\sim
\left(\frac{3\chi\,\eta\,a\,f}{4\pi(\rho_p-\rho_w)g}\right)^{1/2}
\sim f^{1/2},
\end{equation}
where $\rho_p$ and $\rho_w$ are the particle and fluid densities, respectively.
The square-root scaling follows directly from the lift scaling linearly with $f$ and decaying as $r^{-2}$, consistent with the exponent measured in Fig.~\ref{fig2_new}d. Simulations provide further support for this mechanism, directly showing the particle moving away from the obstacle while rising above the substrate before escaping (Fig.~S1). For much larger obstacles, this asymmetry is suppressed, consistent with the stable hydrodynamic trapping reported previously~\cite{van_der_wee_simple_2023}.

\subsection{Penetration and scattering at the obstacle lattice edge}

\begin{figure}[!ht]
 \begin{center}
 \includegraphics[width=1\linewidth]{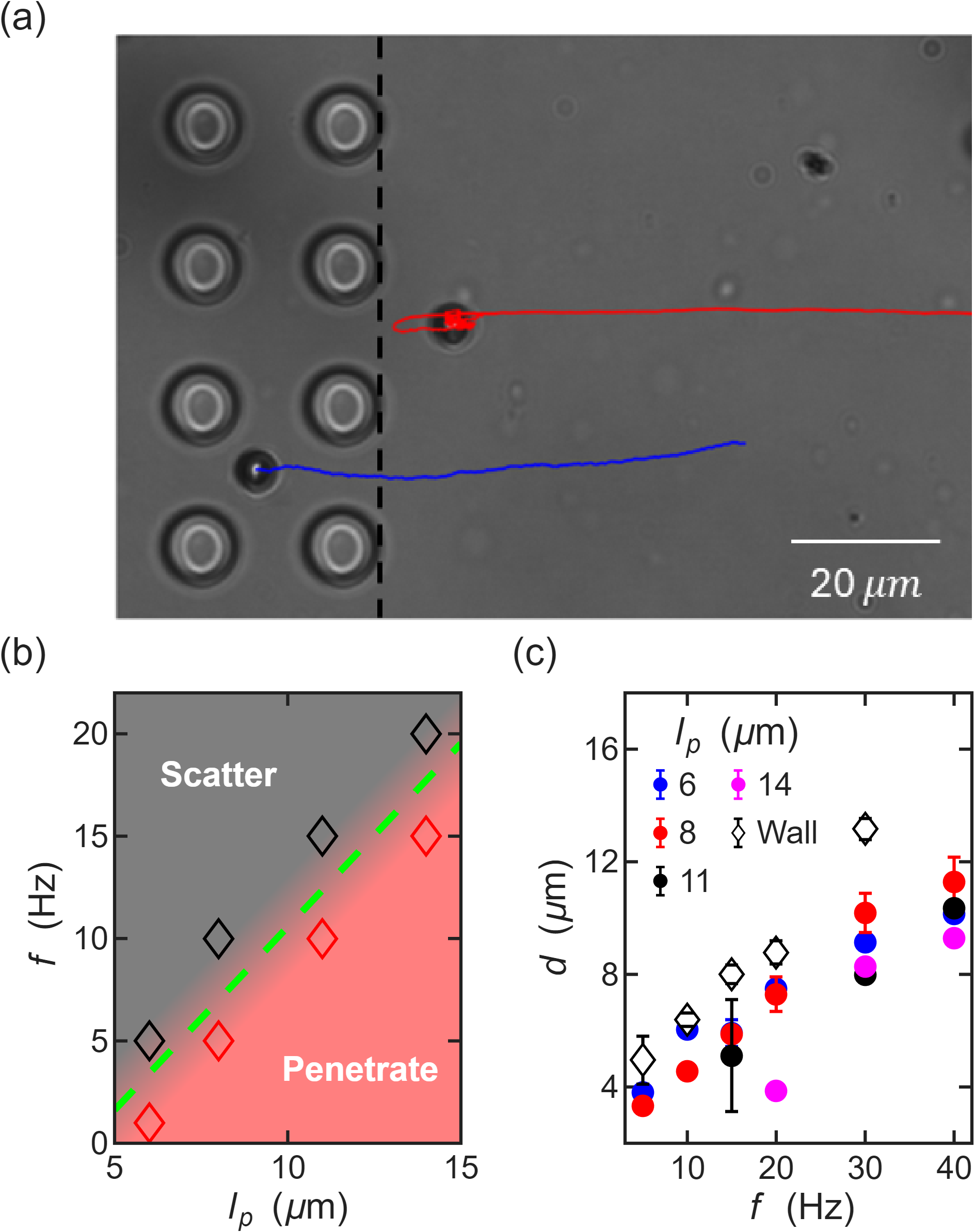}
 \end{center}
\caption{\textbf{Penetration and scattering at the obstacle lattice edge.}
\textbf{(a)} Representative trajectories at normal incidence, $\theta = 90^\circ$, for two rotation frequencies. At low frequencies, the particle penetrates the lattice; at higher frequencies, it is scattered and remains at a steady-state distance $d$ from the edge.
\textbf{(b)} State diagram in the ($l_p$, $f$) plane. Red diamonds mark the highest frequency at which penetration is observed for each lattice spacing; black diamonds mark the lowest frequency at which scattering is observed. The transition shifts to higher frequencies as $l_p$ increases, consistent with a critical frequency $f^* \sim l_p^2$. Green dashed line represents the fit of the midpoints between scattering and penetration.
\textbf{(c)} Steady-state distance $d$ as a function of rotation frequency for different lattice spacings. Open diamonds correspond to a continuous wall.}
 \label{fgr}
\end{figure}

 \begin{figure}[!ht]
\centering
\includegraphics[width=0.95\linewidth]{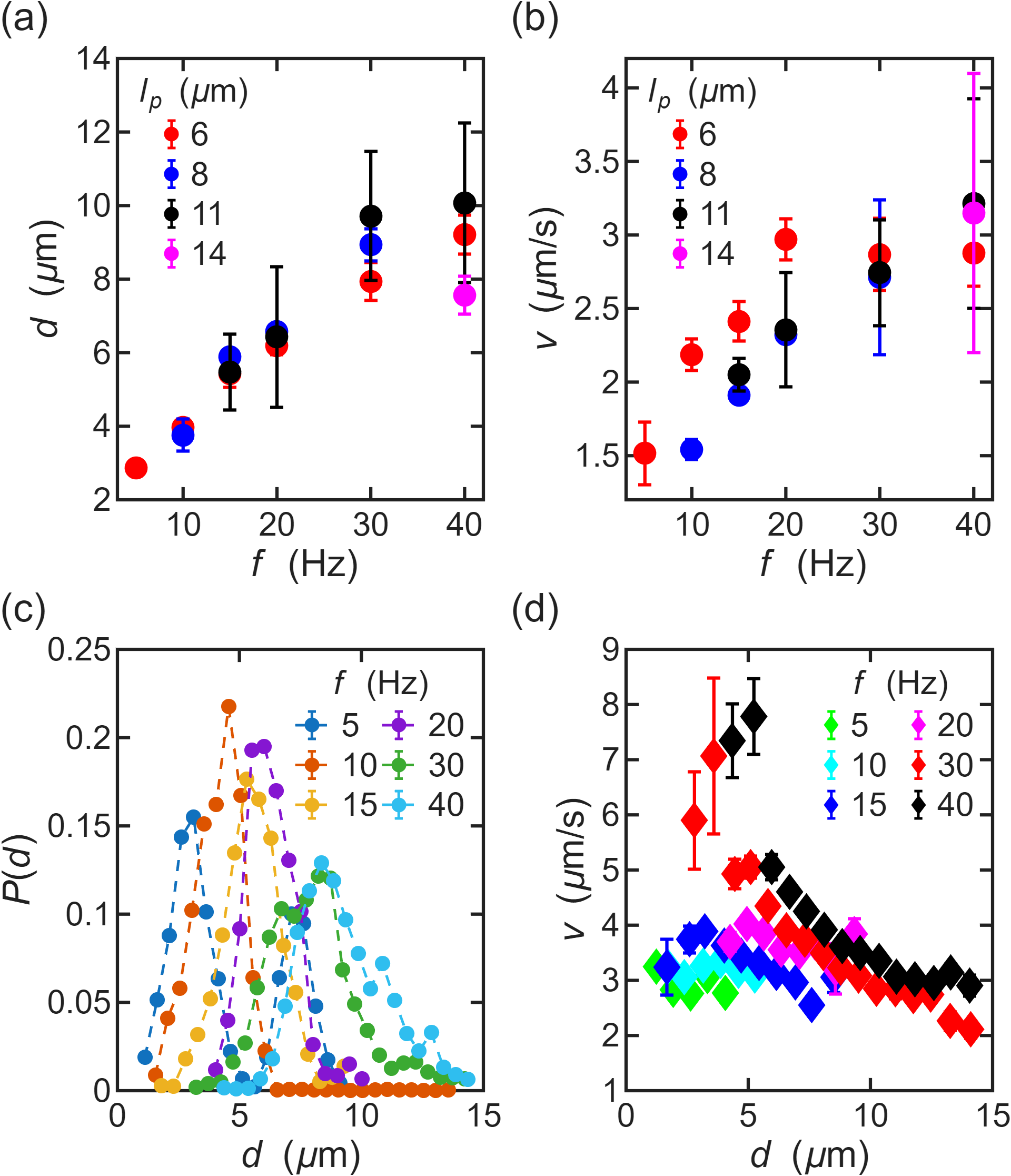}
\caption{\textbf{Sliding dynamics along a tilted lattice edge.}
\textbf{(a)} Steady-state distance from the lattice edge, $d$, as a function of $f$, for different lattice spacings $l_p$.
\textbf{(b)} Sliding velocity $v_\mathrm{sc}$ parallel to the lattice edge as a function of $f$. For large $l_p$, $v_\mathrm{sc}$ increases with $f$, whereas for the densest lattice ($l_p = 6~\mu\mathrm{m}$) it saturates, reflecting the coupling between the rotation frequency and the distance from the edge.
\textbf{(c)} Probability distribution of the steady-state distance $d$ for $l_p = 6~\mu\mathrm{m}$ at different $f$. The distribution shifts to larger $d$ and broadens with increasing frequency, indicating weaker confinement farther from the edge.
\textbf{(d)} Sliding velocity as a function of $d$ for $l_p = 6~\mu\mathrm{m}$. The non-monotonic dependence marks the crossover from discrete obstacle-mediated to wall-like sliding.}
\label{fig:sc-angle}
\end{figure}

Arranging the obstacles in a periodic lattice introduces a new length scale, the lattice spacing $l_p$, defined as the gap between neighboring obstacles. We therefore hypothesize that the transport regime is controlled by the competition between the obstacle spacing $l_p$ and the single-obstacle scattering distance $r_\mathrm{sc}$. When $r_\mathrm{sc} < l_p/2$, the scattering regions of neighboring obstacles remain separated and the particle can penetrate the lattice. By contrast, when $r_\mathrm{sc} \gtrsim l_p/2$, these regions overlap and form a continuous hydrodynamic barrier that redirects the particle along the lattice edge. Since $r_\mathrm{sc} \sim f^{1/2}$, this criterion predicts a critical frequency $f^* \sim l_p^2$ separating penetration from scattering. We test this prediction by driving particles at normal incidence ($\theta = 90^\circ$) toward square obstacle lattices while varying $l_p$ and $f$. At low frequencies, particles penetrate the lattice, whereas at higher frequencies they are scattered and remain at a steady-state distance $d$ from the edge (Fig.~\ref{fgr}a), as also reproduced by numerical simulations (see Movie SM3).
The state diagram in Fig.~\ref{fgr}b maps the boundary between these regimes across the explored $(l_p, f)$ parameter space. The transition shifts to higher frequencies as $l_p$ increases, in agreement with the predicted scaling $f^* \sim l_p^2$.

The steady-state distance $d$ (Fig.~\ref{fgr}c) increases with $f$ for all lattice spacings, following the same scaling $d \sim f^{1/2}$ as the single-obstacle scattering distance $r_\mathrm{sc}$ (ESI, Fig.~S3). The scattering mechanism identified for a single obstacle therefore persists when the obstacles are arranged in a lattice: when the scattering regions of neighboring obstacles overlap, the lattice edge acts as a hydrodynamic barrier. The same scaling is observed for a continuous wall, for which $d$ is larger than for the obstacle lattices at the same frequency.

The penetration-to-scattering transition also suggests a possible route for separating particles with different physical properties under a common external field. From the balance between hydrodynamic lift and buoyant weight, the scattering distance depends not only on the imposed rotation frequency but also on the particle density. At a fixed driving frequency, particles of the same size but different densities can therefore exhibit different values of $r_\mathrm{sc}$. A second route exploits the magnetic response of the particles. Above a step-out frequency set by the particle magnetic moment and the field amplitude, particles no longer rotate synchronously with the field and instead rotate at a lower mean rate. Particles with different magnetic properties can therefore rotate at different mean frequencies under the same field, and acquire different scattering distances. In both cases, for a suitable obstacle spacing, a particle with $r_\mathrm{sc}<l_p/2$ can penetrate the lattice, whereas one with $r_\mathrm{sc}\gtrsim l_p/2$ is scattered and redirected along its edge. The obstacle spacing and the externally imposed field therefore set the operating conditions of the device, while differences in particle properties determine the resultant transport pathways. Graded obstacle lattices, in which variations in $l_p$ translate differences in scattering distance into distinct particle trajectories, could further enable spatial separation.

\subsection{Guided sliding along a tilted lattice edge}

\begin{figure*}[t]
\centering
\includegraphics[width=0.95\linewidth]{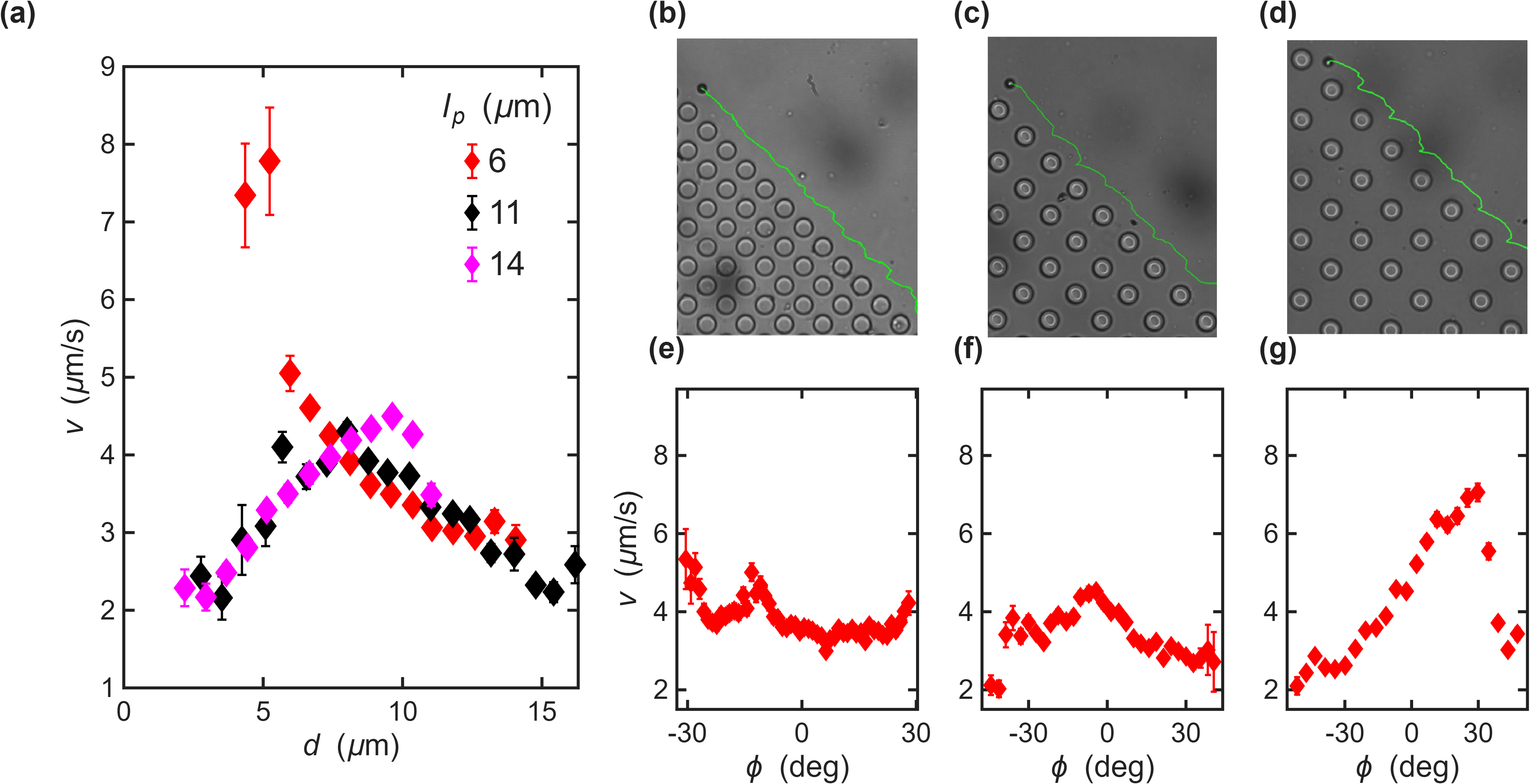}
\caption{\textbf{Crossover from wall-like guidance to discrete obstacle-to-obstacle motion.}
\textbf{(a)} Sliding velocity $v_\mathrm{sc}$ as a function of steady distance $d$ for different lattice spacings $l_p$. The velocity maximum shifts to larger $d$ as $l_p$ increases.
(\textbf{b,c,d}) Representative sliding trajectories for $l_p = 6$, 11, and 14~$\mu\mathrm{m}$, respectively. The local angle $\phi$ is the angle between $\hat{\mathbf{e}}_0$ and the vector from the nearest obstacle center to the particle.
(\textbf{e,f,g}) Sliding velocity $v_\mathrm{sc}$ as a function of $\phi$ for the same lattice spacings. The angular modulation grows with $l_p$, marking the transition from smooth to discrete sliding.
}
\label{fig:edge_velocity}
\end{figure*}

Tilting the lattice edge transforms scattering into directed transport. At an incidence angle $\theta = 45^\circ$, particles still penetrate the lattice below a critical frequency and scatter above it, but the scattered state acquires a component of motion parallel to the boundary. Because the obstacle-induced pressure arrests penetration without suppressing the tangential component of the rolling motion, the particle enters a sustained sliding state at a nearly constant distance $d$ from the lattice edge, with velocity $v_\mathrm{sc}$ directed parallel to the edge (see Movies SM4 and SM5). The lattice edge thus acts as a hydrodynamic guide, converting scattering into persistent motion along its length. Varying the incidence angle, we find that $d$ increases and $v_\mathrm{sc}$ decreases with $\theta$ (ESI, Fig.~S4). The steady sliding distance $d$ increases with rotation frequency for all lattice spacings examined (Fig.~\ref{fig:sc-angle}a), consistent with the lift mechanism identified for single-obstacle scattering. The sliding velocity $v_\mathrm{sc}$ shows a different dependence. For large obstacle spacings, $v_\mathrm{sc}$ increases with $f$, whereas for the densest lattice ($l_p = 6~\mu\mathrm{m}$) it increases at low frequencies and then saturates (Fig.~\ref{fig:sc-angle}b). Increasing the rotation frequency therefore has two competing consequences: it increases the rolling velocity while simultaneously displacing the particle farther from the lattice edge.

This competition is reflected in the non-monotonic dependence of $v_\mathrm{sc}$ on $d$ (Fig.~\ref{fig:sc-angle}d). At small $d$, the hydrodynamic interaction is strongly modulated by the discrete obstacle lattice, and the pressure fields generated by nearby obstacles hinder motion along the edge. Increasing $d$ reduces the influence of these local pressure variations and therefore increases $v_\mathrm{sc}$. At larger distances, the particle becomes less sensitive to individual obstacles and the lattice row behaves increasingly as an effective wall. Because the hydrodynamic coupling to the obstacle row weakens with distance, $v_\mathrm{sc}$ decreases as $d$ increases further, producing a maximum at an intermediate separation. For the densest lattice ($l_p = 6~\mu\mathrm{m}$), increasing $f$ shifts the distribution of steady-state distances toward larger $d$ and broadens it (Fig.~\ref{fig:sc-angle}c). The broader distribution indicates greater variability in the selected sliding distance, which may reflect variations in the local hydrodynamic interactions between the particle and the obstacle row. Together with the non-monotonic dependence of $v_\mathrm{sc}$ on $d$, this provides an explanation for the saturation of the sliding velocity at high frequencies. Increasing $f$ enhances the intrinsic rolling velocity but also pushes the particle towards larger distances, eventually beyond the separation at which sliding is fastest. These competing effects limit the increase of $v_\mathrm{sc}$ with driving frequency.

This crossover from discrete to wall-like sliding is further revealed by the dependence of $v_\mathrm{sc}$ on $d$ and on the angular coordinate $\phi$ across different obstacle spacings (Fig.~\ref{fig:edge_velocity}a--g). Here, $\phi$ is defined as the angle between $\hat{\mathbf{e}}_0$ and the vector joining the nearest obstacle center to the particle center. For closely spaced obstacles, the variation of $v_\mathrm{sc}$ with $\phi$ is weak, indicating that the hydrodynamic interaction varies only weakly along the lattice edge. As $l_p$ increases, the angular modulation becomes more pronounced, reflecting the increasing influence of individual obstacles on the sliding trajectory. The velocity maximum shifts to larger $d$ as $l_p$ increases (Fig.~\ref{fig:edge_velocity}a), so that, at the frequency range examined, the particle in a denser lattice is pushed beyond the maximum, whereas in a more open lattice it is not.

\subsection{Collective sliding along the lattice edge}

\begin{figure}[h!]
\centering
\includegraphics[width=0.95\linewidth]{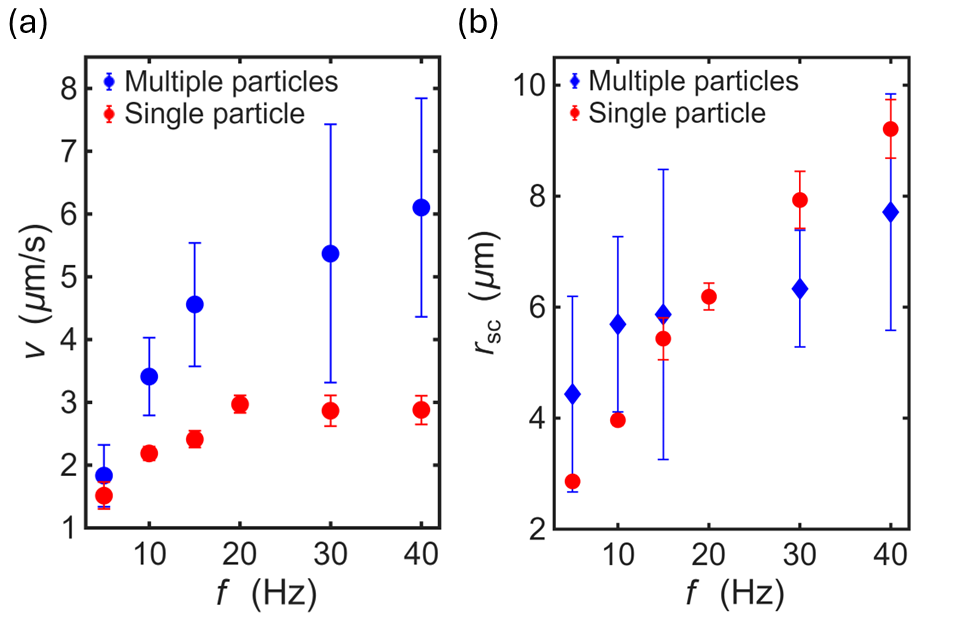}
\caption{\textbf{Collective sliding of multiple microrollers along the lattice edge.}
\textbf{(a)} Sliding velocity $v_\mathrm{sc}$ as a function of rotation frequency for multiple particles (blue) and a single particle (red). Multiple particles slide faster than a single particle across the full frequency range, with the enhancement increasing with $f$.
\textbf{(b)} Steady-state distance $d$ as a function of rotation frequency for multiple particles (blue) and a single particle (red). The comparable values indicate that the velocity enhancement does not arise primarily from a change in the distance from the edge.}
\label{fig:collective}
\end{figure}

When several microrollers slide along the lattice edge simultaneously, their motion becomes collective. In addition to the hydrodynamic interaction with the obstacle row that guides an isolated particle, each roller generates a flow field that acts on its neighbors. Particles sliding together move faster than an isolated particle across the full frequency range examined (Fig.~\ref{fig:collective}a, $l_p = 6~\mu\mathrm{m}$), with the enhancement increasing with $f$ and reaching approximately a factor of two at the highest frequencies. By contrast, their steady distance $d$ from the lattice edge remains comparable to that of an isolated particle within experimental uncertainty, although with a broader distribution (Fig.~\ref{fig:collective}b). The enhanced velocity therefore cannot be attributed primarily to a change in particle--edge separation and instead points to hydrodynamic interactions between neighboring rollers that assist their motion along the boundary, analogous to the cooperative motion of driven colloids in one-dimensional environments~\cite{ketzetzi2022activity,martinez2015colloidal}.

The particles organize into train-like configurations, in which neighboring rollers move with
similar velocities while maintaining unequal and fluctuating separations (see Movie SM6). These variations in the instantaneous particle arrangement provide a natural origin for the broader distribution of $d$ and the larger velocity fluctuations observed during collective sliding (Fig.~\ref{fig:collective}). The obstacle row confines the rollers to a common path, allowing particle--particle hydrodynamic interactions to act persistently along the direction of propagation. These results show that the lattice edge not only guides individual particles but also supports hydrodynamically coupled collective transport with velocities exceeding those of isolated rollers. 

\section{Conclusions}\label{sec13}

We have shown that the transport of driven microrollers through obstacle lattices is governed by two distinct geometrical effects. First, finite obstacle height breaks the vertical symmetry of the rotational flow, generating a hydrodynamic lift that weakens substrate-induced shear and gives rise to a frequency-dependent scattering distance, $r_\mathrm{sc}\sim f^{1/2}$. The competition between this distance and the obstacle spacing determines whether particles penetrate the lattice or are scattered along its boundary. Second, the lateral discreteness of the obstacle row shapes the subsequent edge-guided motion, producing a non-monotonic dependence of the sliding velocity on particle--edge separation. The same boundary also supports collective sliding, with hydrodynamically coupled particles moving faster than isolated rollers. More broadly, our results illustrate how finite-size boundaries can introduce driving-frequency-dependent transport regimes even at low Reynolds numbers. Although hydrodynamic forces scale linearly with the rotation frequency, the balance between obstacle-induced lift and the frequency-independent buoyant weight introduces an additional length scale that governs scattering. The interplay between obstacle geometry and driving frequency therefore provides a physical basis for controlling particle trajectories in structured environments, with potential applications in particle separation and edge-guided transport.

\section*{Author contributions}
J.L.A. and L.R.A contributed to conceptualization, investigation, resources, writing, supervision, funding acquisition, while G.G. (experiments) carried out investigation (experiments), formal analysis, writing - original draft, and A.E.O performed investigation (simulations) and formal analysis. 

\section*{Conflicts of interest}
There are no conflicts to declare.

\section*{Data availability}
The numerical data underlying the figures in the
main text and Supplementary Information are
openly available in the Zenodo repository at
\url{https://doi.org/10.5281/zenodo.24563305}.

\section*{Acknowledgements}
The authors acknowledge financial support by
MCIN/AEI/10.13039/501100011033/ for all grants
listed next: JLA and LRA for PID2022-143010NB-I00 and CEX2023-001316-M, also supported by ``ERDF A way of making Europe''; GG for PRE-2021-099492, also supported by ``ESF Investing in your future''. The authors thank Prof. A. Hidalgo for providing access to the plasma etcher. The authors thank Rodrigo Leodan Espinoza Correa for fabrication of the master templates. The authors thank Profs. J.V. Alvarez and E. Lauga for insightful discussions.



\end{document}